# LEARNING-BASED PERSONAL SPEECH ENHANCEMENT FOR TELECONFERENCING BY EXPLOITING SPATIAL-SPECTRAL FEATURES

*Yicheng Hsu[1], Yonghan Lee[1], and Mingsian R. Bai[1,2]*

[1]Department of Power Mechanical Engineering, National Tsing Hua University, Taiwan
[2]Department of Electrical Engineering, National Tsing Hua University, Taiwan
shane.ychsu@gapp.nthu.edu.tw, yong-han@gapp.nthu.edu.tw, msbai@pme.nthu.edu.tw

## ABSTRACT

Teleconferencing is becoming essential during the COVID-19 pandemic. However, in real-world applications, speech quality can deteriorate due to, for example, background interference, noise, or reverberation. To solve this problem, target speech extraction from the mixture signals can be performed with the aid of the user's vocal features. Various features are accounted for in this study's proposed system, including speaker embeddings derived from user enrollment and a novel long-short-term spatial coherence (LSTSC) feature pertaining to the target speaker activity. As a learning-based approach, a target speech sifting network was employed to extract the target signal. The network trained with LSTSC in the proposed approach is robust to microphone array geometries and the number of microphones. Furthermore, the proposed enhancement system was compared with a baseline system with speaker embeddings and interchannel phase difference. The results demonstrated the superior performance of the proposed system over the baseline in enhancement performance and robustness.

## 1. INTRODUCTION

Teleconferencing has become essential during the COVID-19 pandemic. In teleconferencing, the enhancement of target speech signals in the presence of competing speech (e.g., interfering dialogue from a television) remains a challenge. This issue can be addressed using a speech separation system that serves as the frontend for speech signal processing. The mixed signal is separated into independent continuous streams associated with each speaker, and the target speaker is then identified from among the separated signals. Many techniques have been suggested for performing source separation in the time–frequency (TF) domain [1, 2, 3, 4, 5]. However, these separation methods are difficult to apply in real-world scenarios when prior knowledge of the number of sources is unavailable. In stay-at-home applications, regardless of the source relevance, the aforementioned comprehensive separation approach can be superfluous and computationally demanding because, in general, only one source is of interest.

As opposed to the aforementioned separation approaches, a target-oriented enhancement approach is proposed in this paper based on the technique reported in [6]. In the proposed approach, only the speech signal of the target speaker is extracted by exploiting the auxiliary information of the target speaker. The auxiliary information can be obtained from pre-enrolled utterances of the target speaker [7, 8, 9, 10, 11], video imagery of the target speaker [12, 13, 14, 15], electroencephalogram signals [16], and speech activities of the target speaker [17]. Although these approaches (mostly single-channel ones) are efficient in extracting close-talking target speech, the enhancement performance often degrades due to interference, reverberation, noise, etc. In the face of these adverse conditions, multichannel approaches can be more advantageous compared to single-microphone approaches as these leverage the spatial information provided by a microphone array.

With additional spatial cues, improved enhancement performance can be achieved. For instance, the direction-aware SpeakerBeam [18] combines an attention mechanism with beamforming. The neural spatial filter [19] uses the directional information while extracting target speech. The time-domain SpeakerBeam (TD-SpeakerBeam) [20] employs interchannel phase differences (IPDs) as additional input features to increase the speaker separation capability. Instead of *ad hoc* spatial features [21, 22], better spatial feature estimates can be obtained from multichannel microphone signals by using a trainable spatial encoder. Inspired by enhancement techniques employing the aforementioned spatial cues, in this paper, we proposed a learning-based personal speech enhancement system by exploiting spatial spectral features. The proposed spatial feature was partly inspired by the work of Laufer-Goldshtein et al. [23] who employed RTFs for voice activity detection.

In this study, we developed a personal speech enhancement system based on a convolution recurrent network (CRN) model with spatial features and pre-enrolled utterances as inputs. A novel long-short-term spatial coherence (LSTSC) feature is computed as spatial features in relation to the speaker activity pertaining to each time–frequency bin. With LSTSC as the input features, the spatially varying source signal are extracted by using the decoder layers. However, only one of the sifted sources (i.e., the target speaker) is of interest; thus, speaker-dependent information is needed to further extract the target speaker signals from the mixture. The generalization of a network model to array geometries and channel count other than the array configurations used in the training phase is difficult. The results revealed that the proposed system is robust to alteration of array configurations. To our knowledge, this is the first attempt at speech enhancement using the proposed LSTSC. To assess the robustness of the proposed approach, two sets of simulations were conducted: the first was conducted to compared the improvement in performance attained using the proposed LSTSC feature versus a conventional IPD feature when unseen array geometries are involved in the testing phase, and the second was conducted to determine the influence of the number of microphones, including the single-microphone case where no spatial information is available. In the simulation, short-time objective intelligibility (STOI) [24] and perceptual evaluation of speech quality (PESQ) [25] were employed as performance metrics.

## 2. PERSONAL SPEECH ENHANCEMENT SYSTEM

In this section, we describe a system aimed at extracting the target speech signal from the multichannel mixture by using pre-enrolled utterances of the target speaker. Here, we consider a scenario in which utterances by a target speaker are captured using distant microphones in the presence of speech-like interferences from a television placed nearby. Therefore, the interference represents a source type that is persistent throughout time and fixed in space. No prior knowledge of the array configuration and the directions of the target speaker and the interferences is available. Fig. 1 illustrates the block diagram of the proposed personal speech enhancement system comprising three key modules: 1) the spatial feature extraction module (Section 2.1), which exploits LSTSC as spatial cues; 2) the speaker encoder network (Section 2.2), which produces a speaker embedding from the pre-enrolled utterances; and 3) the target-speech-sifting CRN (Section 2.3) (including an encoder, an LSTM layer, and a decoder), which extracts the target speech signal based on the auxiliary information provided by the two preceding modules.

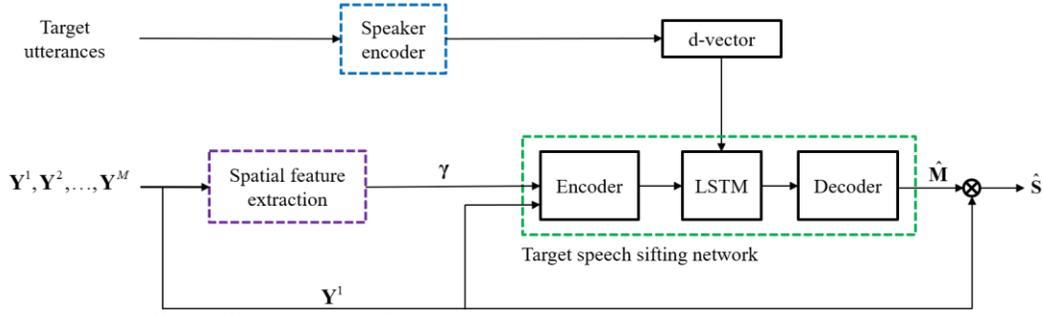

**Fig. 1**. Block diagram of the proposed personal speech enhancement system.

## 2.1. Spatial feature extraction

Consider one static interference source and one target speaker in a reverberant room. The signals emitted by the sources are captured by a microphone array containing $M$ elements and are analyzed in the short-time Fourier transform (STFT) domain. Suppose that the target speaker and a spatially stationary interference source coexist in a room. The problem can be formulated in the STFT domain, with $l$ denoting the time index and $f$ denoting the frequency index. The signal captured by the $m$th microphone can be written as

$$Y^m(l,f) = \sum_{j=1}^{J} A_j^m(f) S_j(l,f) + V^m(l,f). \tag{1}$$

where $m \in \{1,\ldots,M\}$ denotes a given microphone, $l \in \{1,\ldots,T\}$ denotes the time frame, $f \in \{1,\ldots,F\}$ denotes the frequency bin, $Y_j^m(l,f) = A_j^m(f) S_j(l,f)$ is the signal of the $j$th source measured by the $m$th microphone, $A_j^m(f)$ is the acoustic transfer function (ATF) relating the $j$th source and the $m$th microphone, $S_j(l,f)$ is the signal of the $j$th source, and $V^m(l,f)$ is the nondirectional noise measured by the $m$th microphone.

For each TF bin, the short-term relative transfer function (RTF) between the $m$th microphone and reference microphone 1 can be estimated by averaging $(R + 1)$ frames:

$$\tilde{R}^m(l,f) \equiv \frac{\hat{\Phi}_{y^m y^1}}{\hat{\Phi}_{y^1 y^1}} \equiv \frac{\sum_{n=l-R/2}^{l+R/2} Y^m(n,f) Y^{1*}(n,f)}{\sum_{n=l-R/2}^{l+R/2} Y^1(n,f) Y^{1*}(n,f)} \tag{2}$$

where * denotes the complex conjugate operation, $\hat{\Phi}_{y^m y^1}$ is the short-time cross-spectral density estimate between channels $m$ and 1, and $\hat{\Phi}_{y^1 y^1}$ is the short-time autospectral density of the reference microphone.

As a key step, a "whitened" feature vector $\mathbf{r}(l,f) \in \mathbb{R}^{M-1}$ pertaining to each TF bin is calculated as follows:

$$\mathbf{r}(l,f) = \left[ \frac{\tilde{R}^2(l,f)}{|\tilde{R}^2(l,f)|}, \ldots, \frac{\tilde{R}^M(l,f)}{|\tilde{R}^M(l,f)|} \right]^T \tag{3}$$

where $|\cdot|$ is the complex modulus.

For spatially stationary sources, the following long-term RTF computed via recursive averaging is used to approximate the expectation of time-average of the feature vector:

$$\bar{r}^m(l,f) = \lambda \bar{r}^m(l-1,f) + (1-\lambda) r^m(l,f), \quad m = 2,\ldots,M, \tag{4}$$

where $\lambda$ is the forgetting factor that facilitates the tuning between the global view (large $\lambda$) and the local view (small $\lambda$) of time frames. The feature vector $\bar{\mathbf{r}}(l,f)$ is also whitened after each recursive step:

$$\bar{\mathbf{r}}(l,f) = \left[ \frac{\bar{r}^2(l,f)}{|\bar{r}^2(l,f)|}, \ldots, \frac{\bar{r}^M(l,f)}{|\bar{r}^M(l,f)|} \right]^T. \tag{5}$$

To fully exploit the temporal–spatial information conveyed by the whitened RTF, we propose to calculate the LSTSC, $\gamma_{lf}(l,f)$, between the short-term whitened feature vector $\mathbf{r}(l,f)$ and the long-term whitened feature vector $\bar{\mathbf{r}}(l,f)$:

$$\gamma(l,f) \approx \frac{1}{M-1} \sum_{m=2}^{M} \frac{\text{Re}\{\tilde{R}^m(l,f) \bar{r}^m(l,f)^*\}}{|\tilde{R}^m(l,f)||\bar{r}^m(l,f)|} \approx \frac{1}{M-1} \text{Re}\{\mathbf{r}^H(l,f) \bar{\mathbf{r}}(l,f)\}, \tag{6}$$

where $\text{Re}\{\cdot\}$ denotes the real-part operator. The Euclidean angle [26] rather than the Hermitian angle is adopted in the LSTSC definition to ensure sign sensitivity. The LSTSC is an indicator of spatial correlation between the short-term RTF and the long-term RTF (associated with the spatially stationary interference). In addition, $-1 \leq \gamma(l,f) \leq 1$. The larger the LSTSC feature (close to 1), the more likely the current TF bin pertains to the interference. Otherwise, for LSTSC < 1 it is likely for either the target speaker to be present or silence period to occur. As indicated in Fig. 2, LSTSC values with a large $\lambda$ (global LSTSC) are used to sift out the TF bins that either correspond to the active target or correspond to both the target and the interference (which are rendered inactive), whereas LSTSC values with a small $\lambda$ (local LSTSC) are used to identify the TF bins corresponding to the directional sources.

## 2.2. Speaker encoder

The speaker encoder generates a speaker embedding from the pre-enrolled utterances of the target speaker. A speaker embedding can be extracted using a speaker encoder that is jointly trained using the target speech model or a pretrained model for extraction of speaker information, such as the i-vector [27], x-vector [28], or d-vector [29]. The target speaker utterances are pre-recorded using an enrollment process. In this study, we used the d-vector, which has been successfully used in various applications such as speaker diarization, speech synthesis, speech translation, personal voice activity detection, source separation, and voice preservation tests. The proposed model comprising a three-layer LSTM network was trained using the generalized end-to-end loss function suggested in [29]. The speaker encoder was trained using the VoxCeleb2 data set [30]. The model yields embeddings in sliding windows. The resulting aggregated embedding, known as the d-vector, encodes the voice characteristics of the target speaker.

## 2.3. Target speech sifting network

The target-speech-sifting network is based on a CRN. As illustrated in Fig. 1, the network has three inputs: LSTSC calculated from the microphone signals, the d-vector of the target speaker generated by the speaker encoder, and the magnitude spectrogram computed for the reference microphone. In this network, the spatially stationary persistent interference is suppressed by the encoder layers. Then, the d-vector is repeatedly concatenated to the output of the previous layer of the encoder in every time frame. The resulting vector is then fed to the following LSTM layers to sift the latent features pertaining to the target speaker. The soft mask generated by the network is multiplied element-wise with the noisy magnitude spectrogram to yield an enhanced spectrogram. The complete complex spectrogram can be obtained by combining the enhanced magnitude spectrogram with the phase of the noisy spectrogram. The enhanced time-domain signals can be obtained using inverse STFT and the overlap-add method. The network is trained to minimize the mean square error between the masked magnitude spectrogram and the target magnitude spectrogram of the clean audio.

Table 1 summarizes the parameters of the proposed target-speech-sifting network architecture. The input and output size of each layer are given in the (featureMaps × timeSteps × frequencyChannels) format, and the layer hyperparameters are given in the (kernelSize, strides, outChannels) format. To ensure stable convergence property of the network, skip connections are used to concatenate the output of each encoder layer and the input of the corresponding decoder layer. In the CRN, all convolution and deconvolution operations are causal; thus, no future information is needed for mask estimation at each time frame. The number of feature maps in each decoder layer is doubled in size because of the skip connections. Exponential linear units (ELUs) are employed in all convolution and deconvolution layers, except for the output layer, where sigmoid activation is used for mask estimation.

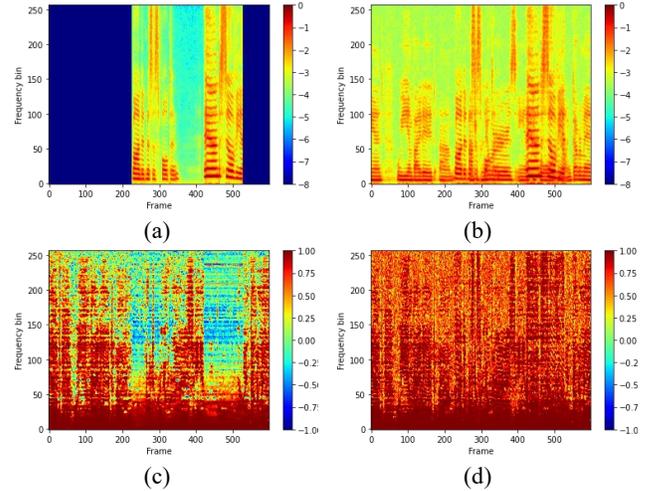

**Fig. 2**. Example spectrogram of the LSTSC feature calculated with different forgetting factors: (a) target speech signal, (b) noisy mixture signal, (c) global LSTSC ($\lambda$ = 0.999), and (d) local LSTSC ($\lambda$ = 0.1).

**Table 1**. Parameters of the proposed target speech sifting network.

| Layer | Input size | Hyperparameters | Output size |
|---|---|---|---|
| conv2d_1 | 2 × T × 257 | 1 × 3, (1, 2), 4 | 4 × T × 128 |
| conv2d_2 | 4 × T × 128 | 1 × 3, (1, 2), 8 | 8 × T × 63 |
| conv2d_3 | 8 × T × 63 | 1 × 3, (1, 2), 16 | 16 × T × 31 |
| conv2d_4 | 16 × T × 31 | 1 × 3, (1, 2), 32 | 32 × T × 15 |
| conv2d_5 | 32 × T × 15 | 1 × 3, (1, 2), 64 | 64 × T × 7 |
| conv2d_6 | 64 × T × 7 | 1 × 3, (1, 2), 128 | 128 × T × 3 |
| reshape_1 | 128 × T × 3 | - | T × 384 |
| lstm | T × (384+256) | 384 | T × 384 |
| reshape_2 | T × 384 | - | 256 × T × 3 |
| deconv2d_6 | 256 × T × 3 | 1 × 3, (1, 2), 64 | 64 × T × 7 |
| deconv2d_5 | 128 × T × 7 | 1 × 3, (1, 2), 32 | 32 × T × 15 |
| deconv2d_4 | 64 × T × 15 | 1 × 3, (1, 2), 16 | 16 × T × 31 |
| deconv2d_3 | 32 × T × 31 | 1 × 3, (1, 2), 8 | 8 × T × 63 |
| deconv2d_2 | 16 × T × 63 | 1 × 3, (1, 2), 4 | 4 × T × 128 |
| deconv2d_1 | 8 × T × 128 | 1 × 3, (1, 2), 1 | 1 × T × 257 |

## 3. SIMULATIONS

### 3.1 Data preparation

To validate the proposed personal speech enhancement system, we used data from two open data sets. The clean utterances for training and testing were selected from the train-clean-360 and dev-clean subsets of the LibriSpeech corpus [31], which contains utterances from 921 and 40 speakers, respectively. We generated noisy training and testing data using the VoxConverse data set [32], from which 74-h human conversation clips from YouTube were chosen. The audio contained noise of various types, such as background noise, music, laughter, and applause. The simulation is conducted in a sample rate of 16 kHz.

In the training phase, a reference speech signal different from the clean signal is randomly selected for enrollment among all the utterances of the target speaker. Noisy audio signals in the form of 6-s clips were prepared by mixing clean target speech signal and speech-like interference for signal-to-noise ratios (SNRs) of 0, 5, 10, and 15 dB. In the training and validation sets, a four-element uniform circular array (UCA) with a radius of 3.5 cm was employed. In the testing set, three array configurations were employed to evaluate the robustness of the proposed enhancement system, including a four-element UCA with the same radius (3.5 cm) as in the training set, a four-element UCA with a radius of 7 cm, and a four-element uniform linear array (ULA) with an interelement spacing of 2 cm. The reverberant speech signals received at the microphones were generated by convolving the anechoic clean signals with the respective room impulse responses (RIRs) simulated using the image-source method [33]. In the simulation, the microphone array was placed at the center of a 4 × 4 × 3 $m^3$ room with moderate reverberation (T60 = 0.2 s). The target speaker was placed sequentially on a semicircle of 1-m radius centered at the microphone, from 0° to 180° with 1° increments, whereas the interference was located sequentially on a semicircle of 1.5-m radius centered at the microphone, from 180° to 360° at increments of 1°. In addition, the sample size of the training, validation, and test sets were 20,000, 2000, and 3000, respectively. The STFT settings were a 32-ms window length, a 16-ms hop size, and a 512-point fast Fourier transform. The 257-dimensional magnitude spectrum was calculated from the noisy signal captured by the reference microphone. IPD was used as the baseline spatial feature. The proposed LSTSC feature was calculated based on the short-term RTF in Eq. (2) and averaged using the adjacent frames (R = 2). In this simulation, forgetting factors were selected to be λ = 0.999 and 0.1 for calculating the global LSTSC (G-LSTSC) feature and the local (L-LSTSC) feature.

Table 2. Comparisons of different array geometries in terms of STOI and PESQ.

| | | STOI (in %) | | | | | PESQ | | | | |
|---|---|---|---|---|---|---|---|---|---|---|---|
| | SNR (dB) | 0 | 5 | 10 | 15 | Avg. | 0 | 5 | 10 | 15 | Avg. |
| UCA r=3.5cm | Noisy | 65.65 | 74.46 | 81.34 | 85.79 | 76.81 | 1.42 | 1.65 | 1.91 | 2.22 | 1.79 |
| | IPD | **87.73** | **91.08** | **93.26** | **94.53** | **91.65** | **2.22** | 2.57 | 2.88 | 3.12 | 2.69 |
| | G-LSTSC | 87.08 | 90.65 | 92.95 | 94.26 | 91.24 | 2.17 | 2.53 | 2.85 | 3.10 | 2.66 |
| | GL-LSTSC | 87.57 | 90.98 | 93.19 | 94.46 | 91.55 | 2.21 | **2.57** | **2.89** | **3.14** | **2.70** |
| UCA r=7cm | Noisy | 65.80 | 74.20 | 81.06 | 85.83 | 76.72 | 1.44 | 1.64 | 1.91 | 2.22 | 1.80 |
| | IPD | 74.28 | 79.01 | 82.44 | 85.06 | 80.05 | 1.67 | 1.94 | 2.21 | 2.48 | 2.08 |
| | G-LSTSC | 85.57 | **88.89** | **91.17** | **92.48** | **89.53** | 2.09 | 2.41 | 2.71 | 2.95 | 2.54 |
| | GL-LSTSC | **85.66** | 88.88 | 91.12 | 92.33 | 89.50 | **2.11** | **2.42** | **2.73** | **2.97** | **2.56** |
| ULA d=2.0cm | Noisy | 64.45 | 73.48 | 80.55 | 85.47 | 75.99 | 1.39 | 1.60 | 1.87 | 2.17 | 1.76 |
| | IPD | 10.48 | 10.08 | 12.94 | 18.74 | 13.06 | 1.11 | 1.09 | 1.11 | 1.11 | 1.11 |
| | G-LSTSC | 86.03 | 90.07 | 92.41 | 93.92 | 90.61 | 2.14 | 2.53 | 2.86 | 3.11 | 2.66 |
| | GL-LSTSC | **86.49** | **90.37** | **92.67** | **94.10** | **90.91** | **2.19** | **2.58** | **2.90** | **3.15** | **2.71** |

Table 3. Comparison of enhancement performance in terms of STOI and SDR for different numbers of microphone.

| Metric | STOI (in %) | | | | PESQ | | | |
|---|---|---|---|---|---|---|---|---|
| SNR (dB) | 0 | 5 | 10 | 15 | 0 | 5 | 10 | 15 |
| Noisy | 64.97 | 74.18 | 80.76 | 85.73 | 1.41 | 1.62 | 1.87 | 2.18 |
| 1Mic | 78.26 | 85.37 | 89.58 | 92.39 | 1.68 | 2.03 | 2.36 | 2.69 |
| 2Mics | 81.68 | 86.33 | 89.53 | 91.79 | 1.97 | 2.30 | 2.61 | 2.89 |
| 3Mics | 86.55 | 90.41 | 92.72 | 94.26 | 2.14 | 2.51 | 2.82 | 3.08 |
| 4Mics | **87.22** | **90.90** | **93.06** | **94.51** | **2.18** | **2.54** | **2.85** | **3.11** |

### 3.2 Results

We compared the improvements in performance yielded by the LSTSC feature versus baseline IPD feature. Table 2 reveals that the enhancement system based on GL-LSTSC, which combines the G-LSTSC and the L-LSTSC features, and G-LSTSC and the system based on IPD performed comparably with array geometries that were identical to the training set. The system based on GL-LSTSC slightly outperformed the system based on G-LSTSC. However, if the array geometry was not included in the training data set and if IPD was used as a spatial feature, STOI and PESQ decreased considerably. This result suggests that the enhancement system based on the proposed LSTSC feature is robust against variations in array geometries, which is a desirable property in real-world applications.

Second, we examined the effects of number of microphones on speech enhancement. The IPD-based network trained with a one-array configuration cannot be applied to the test data set gathered using another configuration because the IPD feature is already fixed for a particular array geometry and number of microphones. However, for the proposed LSTSC feature, a change in the number of microphones only results in a change in feature vector dimensions in Eqs. (3) and (5) and not the input dimension in Eq. (6). An additional single-microphone system was also included in the simulation to benchmark the proposed approach. In this case, the network was trained with only the magnitude spectrogram and the d-vector, and no spatial feature was available. As indicated in the results summarized in Table 3, the enhancement performance was significantly improved in terms of STOI and PESQ as the number of microphones was increased for various SNR levels. The single-microphone system, representing the approach with no spatial information, yielded the worst performance, especially for scenarios with low SNR levels.

### 4. CONCLUSIONS

In this study, we developed a speech enhancement system based on spatial coherence and target speaker enrollment. The proposed LSTSC feature is effective in distinguishing the static interfering source and the target speaker in light of the coherence between the long-term whitened RTF and the short-term instantaneous RTF. As opposed to the IPD feature, LSTSC is a spatial feature that is robust to the array geometry and number of microphones used. A DNN-based system was developed to enhance the target speech by leveraging the spatial feature and the d-vector produced by speaker enrollment. In future studies, enhancement techniques will be investigated to address noisy and reverberant speech mixtures recorded in scenarios where multiple sources of interference, competing speakers, and a moving target speaker are involved.

### 5. ACKNOWLEDGMENT

This work was supported by the Ministry of Science and Technology (MOST), Taiwan, under the project number 110-2221-E-007-027-MY3.